\documentclass[conference]{IEEEtran}
\usepackage{amsmath,amssymb}

\usepackage{comment}
\usepackage[amsmath,thmmarks]{ntheorem} 
\usepackage{amssymb}
\usepackage{bm}
\usepackage[dvipdfmx]{graphicx}
\usepackage{color}

\graphicspath{{figure/}}

\theoremstyle{plain}
\theoremseparator{．}
\theoremheaderfont{\normalfont\bfseries}
\theorembodyfont{\upshape}
\newtheorem{dfn}{Definition}
\newtheorem{thm}{Theorem}

\theoremstyle{nonumberplain}
\theoremseparator{}
\theoremheaderfont{\normalfont}
\newtheorem{prf}{(Proof)}

\def\qed{\hfill $\Box$} 
\usepackage{setspace}

\begin{document}
\title{Band Splitting Permutations for Spatially Coupled LDPC Codes Enhancing Burst Erasure Immunity} 
\author{
  \IEEEauthorblockN{Hiroki Mori and Tadashi Wadayama}
  \IEEEauthorblockA{Department Computer Science and Engineering, \\Nagoya Institute of Technology,
    Nagoya,  Japan\\
    Email: mori@it.cs.nitech.ac.jp, wadayama@nitech.ac.jp} 
}

\maketitle
\begin{abstract}
It is well known that spatially coupled (SC) codes with erasure-BP decoding have powerful error correcting capability
over memoryless erasure channels.
However, the decoding performance of SC-codes significantly degrades when they are used over burst erasure channels.
In this paper, we propose band splitting permutations (BSP) suitable for $(l,r,L)$ SC-codes.
The BSP splits a diagonal band in a base matrix into multiple bands in order to enhance the 
span of the stopping sets in the base matrix. As theoretical performance guarantees, 
lower and upper bounds on
the maximal burst correctable length of the permuted $(l,r,L)$ SC-codes are presented.
Those bounds indicate that the maximal correctable burst ratio of the permuted SC-codes is 
given by $\lambda_{max} \simeq 1/k$ where $k=r/l$.  This implies the asymptotic optimality of 
the permuted SC-codes in terms of burst erasure correction.
\end{abstract}

\section{Introduction}

Low-Density Parity-Check (LDPC) codes
that are linear codes defined by extremely sparse parity check 
matrices were developed by Gallager in 1963 \cite{Low-Density Parity-Check}.
The combination of LDPC codes and belief propagation provides 
remarkable error correcting performance with reasonable time complexity. In recent days,
it is easy to find practical applications of LDPC codes in wireless/wired communication systems and
storage systems. Not only a practical importance but also recent theoretical advancement 
produces renewed interests in this field.
Kudekar et. al \cite{spatially coupled code} proposed a new class of LDPC codes,
that is called {\em spatially coupled codes} (SC-codes) and they provided theoretical
arguments on threshold saturation of SC-codes \cite{spatially coupled code}.
The origin of SC-codes is LDPC-convolutional codes that date back to the work due to Felstrom 
and Zigangirov \cite{convolutional code}. Lentmaier et. al \cite{Lentmaier} showed an ensemble
of an LDPC-convolutional code can have  a higher threshold than that of 
a component LDPC code ensemble.
From these works on SC-codes, it is unveiled that well-designed SC-codes have
capacity achieving performance over symmetric memoryless channels.

A burst erasure means a consecutive erased symbols. In many practical situations,
we can observe occurrences of burst erasures due to 
slow fading in mobile wireless communication, buffer overflow at a congested router in a packet based 
network, and media flaw in a magnetic recording system.
A strong erasure correcting code should have high erasure correcting capability not only for memoryless random erasures 
but also for burst erasures.
Ohashi et. al \cite{multi SC} pointed out that SC-codes are not immune to burst erasures 
compared with conventional LDPC codes such as regular LDPC codes.
In a typical decoding process of SC-codes, reliabilities of bit estimation gradually improves 
from both side into inside as a domino toppling. Since a burst erasure interferes the propagation of 
a wave of such reliable estimations, it causes severe degradation on decoding performance.
In order to overcome this difficulties, they proposed a new class of 
multidimensional SC-codes that shows higher immunity against burst erasures.

It is known that the burst erasure correcting capability of LDPC codes depends on a column order 
of parity check matrices of LDPC codes \cite{Wmax}. This is because the minimum length of 
stopping sets determining the burst correcting capability  depends on the column order of a parity check matrix.
In order to enhance the burst erasure correctability,
several heuristic algorithms to improve the column order have been 
prensend by Wadayama \cite{Wadayama column permutation}, Paolini and Chiani \cite{Enrico},
Hosoya et al. \cite{Hosoya column permutation}.
Of course, the column order of a parity check matrix does not affect 
the decoding performance over memoryless erasure channels. 

In this paper, we will propose a class of column permutations that is called 
{\em band splitting permutations} suitable for $(l,r,L)$ SC-codes.
A band splitting permutation is applied to the base matrix of $(l,r,L)$ SC-codes having 
a single diagonal band and
it results in a  column-permuted base matrix with several diagonal bands.
By lifting up the permuted base matrix,
we can obtain a parity check matrix of a permuted $(l,r,L)$ SC-codes.
It will be proved that an appropriate band splitting permutation produces 
permuted $(l,r,L)$ SC-codes that have near optimal minimum length of stopping sets.
The permuted SC-codes constructed in such a way have burst erasure correcting 
superior to those of conventional SC-codes. Upper and lower bounds 
on the minimum length of stopping sets  to be proved in this paper can provide 
theoretical performance guarantees for burst erasure correcting capability of permuted 
SC-codes.

The outline of this paper is as follows.
Section 2 provides notion and fundamental definitions required throughout this paper.
Section 3 presents several theorems regarding stopping sets in a base matrix. %
The band splitting permutations will be defined and analyzed in Section 4.
Results on computer experiments will be shown in Section 5. %

\section{Preliminaries}\label{fundamental matters}

\subsection{$(l,r,L)$ SC-codes}\label{SC}

\begin{figure}[tbp]
\begin{center}
\includegraphics[scale=0.25]{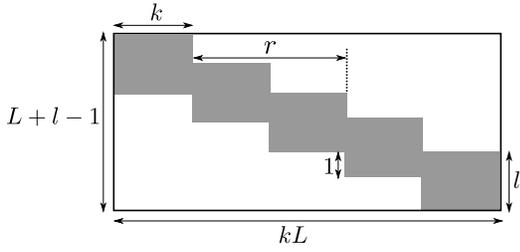}
\end{center}
\caption{Base matrix $B(l,r,L)$ that defines $(l,r,L)$ SC-codes (gray area is filled with symbol one).}
\label{fig:lrL_base_matrix}
\end{figure}

In this subsection, the definition $(l,r,L)$ SC-codes proposed by Kudekar et al. \cite{spatially coupled code}
is reviewed. The $(l,r,L)$ SC-codes belong to the class of protograph LDPC codes 
and its parity check matrix can be obtained by lifting up the base matrix $B(l,r,L)$.
The base matrix $B(l,r,L)$ is a binary $(L+l-1)\times k L$ matrix $(k = r/l)$ and its
structure is illustrated in Fig. \ref{fig:lrL_base_matrix}.
The parameters $l$ and $r$ represents the column weight and maximal row weight of $B(l,r,L)$, respectively.
We assume that the ratio $k=r/l$ is integer throughout the paper.
The parameter $L$ denotes the number of sections. 

A parity check matrix of an $(l,r,L)$ SC-code can be obtained 
by lifting up the base matrix $B(l,r,L)$.
A lift-up process is summarized as follows:
For each element one in $B(l,r,L)$, we can replace it with any binary $M\times M$ permutation matrix. 
The zeros in $B(l,r,L)$ should be replaced with a binary  $M\times M$ zero matrix. 
Let a parity check matrix obtained by the above process be $H$.  The binary linear code defined by $H$
is called an $(l,r,L)$ SC-code. The size of the permutation matrices, $M$, is said to be the {\em lift up factor}.
The number of rows of $H$ is $M (L+l-1)$ 
and the number of columns is $MkL$.
The design rate of $(l,r,L)$ SC-codes, $R(l,r,L)$, is thus given by
\begin{eqnarray}
R(l,r,L)=1-\frac{1}{k}-\frac{l-1}{kL}.
\end{eqnarray}

\subsection{Maximal correctable burst length}\label{Wmax}

Yang and Ryan \cite{Wmax} introduced a measure for burst erasure correcting capability 
of LDPC codes that is called the {\em maximal correctable burst length}. 
Let $H$ be a parity check matrix that defines an LDPC code. 
The maximal correctable burst length of this code is denoted by $W_{max}(H)$.
The meaning of $W_{max}(H)$ is the following.
A burst erasure is a sequence of consecutive erasures occurred on an erasure channel.
In this paper, we assume that only single burst erasure occurs in a code block.
If the length of a single burst erasure is less than or equal to $W_{max}(H)$, it can be 
perfectly corrected by belief propagation (BP) decoding for erasure channels. 
On the other hand, there exists a single burst erasure of length $W_{max}(H)+1$
that cannot be corrected with erasure-BP.
Namely, $W_{max}(H)$ represents the maximum guaranteed correctable length 
for any single burst erasure.
As a related measure for burst erasure correcting capability, we here 
introduce the {\em maximal correctable burst ratio} defined by
$
\lambda_{max}={W_{max}(H)}/{n},
$
where $n$ is the code length. This quantity is useful for studying asymptotic behavior of
the burst correcting capability.

\subsection{Stopping sets and maximal correctable burst length}

\subsubsection{Stooping sets}
Let $H=(\bm{h}_1,\bm{h}_2,\ldots,\bm{h}_n)\in \mathbb{F}_{2}^{m\times n}$ be a parity check matrix.
The vector $\bm{h}_i$ is the $i$-th column vector of $H$.
A sub-matrix of $H$ consists of a subset of column vectors in $H$; namely a sub-matrix of $H$ has the form:
\begin{equation}
H_{\{i_1,\ldots, i_u \}} 
= (\bm{h}_{i_1},\bm{h}_{i_2},\ldots,\bm{h}_{i_u})\in \mathbb{F}_{2}^{m\times u}.
\end{equation}
The subscript in $H_{\{i_1,\ldots, i_u \}}$ represents the column indices of $H$ corresponding to 
the column vectors in the sub-matrix.
\begin{dfn}[Stopping sets \cite{stopping set}]
Let $H$ be a parity check matrix and $S = \{i_1,i_2,\ldots, i_u \} \subseteq [1,n]$ be 
an index set. The notation $[a,b]$ denotes the set of consecutive integers from $a$ to $b$.
 If the sub-matrix $H_S$ has no rows with weight one, the index set $S$ is said to be a stopping set.
\end{dfn}

It is well known that stopping sets are closely related to correctability of erasure patterns 
if we exploit erasure-BP. 
Assume that a transmitted word is a codeword of the code defined by $H$ and
that some symbol erasures happen over the channel. 
Let $E = \{ e_1, e_2, \ldots, e_w \} \subseteq [1,n]$ be the indices corresponding to 
the symbol erasures. This erasure pattern cannot be 
corrected with erasure-BP if there exists a non-empty stopping set $S$ satisfying 
$S \subseteq E$.
This fact indicates that the set of stopping set in $H$ determine $W_{max}(H)$ \cite{stopping set and Wmax}．

Assume that $H=(\bm{h}_1,\bm{h}_2,\ldots,\bm{h}_n)\in \mathbb{F}_{2}^{m\times n}$ is given and
an index set $S = \{i_1, i_2,\ldots, i_u \}\subseteq [1,n]$
is given as well. The length of $S$, that is denoted by $Len(S)$, 
is defined by
\begin{eqnarray}
Len(S)= 1 + \max_{a, b \in S}  |a-b|.
\end{eqnarray}
Let us denote the set of non-empty stopping sets of $H$ by $\mathcal{Q}(H)$.
The {\em span} of $H$, $Span(H)$, is defined by
\begin{eqnarray}\label{eqn:span}
Span(H)=\min_{ S \subset \mathcal{Q}(H)} Len(S).
\end{eqnarray}
It is clear that a burst erasure of length shorter than $Span(H)$
cannot cover any non-empty stopping set in $H$. 
This means that we have
\begin{eqnarray}\label{eqn:Wmax}
W_{max}(H)=Span(H)-1.
\end{eqnarray}
Note that the quantity $W_{max}(H)$ can be evaluated efficiently by using 
erasure-BP \cite{Wmax}. From the definition, we can see that
$Span(H)$ strongly depends on the order of the column vectors in $H$.
It has been shown that an appropriate rearrangement of column order can increase the span of LDPC codes
\cite{Hosoya column permutation}, \cite{Wadayama column permutation}.

\subsubsection{Irreducible stopping sets}

We provide the definition of 
irreducible stopping sets that will be required in the next section.
\begin{dfn}[Irreducible stopping sets]\label{dfn:irreducible}
Let $S \subseteq [1,n]$ be a non-empty stopping set of $H$.
If removing any subset of elements from $S$ yields
an index set that is not a stopping set, then $S$ is said to be irreducible stopping set.
\end{dfn}

From the above definition of irreducible stopping sets, 
it is straightforward to see that the inequality
$
Len(S')\leq Len(S)
$
holds for a pair of nested stopping sets where $S$ is a stopping set and $S' \subseteq S$ is 
an irreducible stopping set in $S$.
From this inequality, we have 
\begin{eqnarray} \label{spandef}
Span(H)=\mbox{min}_{S'\subset \mathcal{Q}'(H)} Len(S'),
\end{eqnarray}
where $\mathcal{Q}'(H)$ is the set of irreducible stopping set of $H$.
This means that we only need to focus on the set of irreducible stopping sets
when we discuss the span of $H$.

\section{Irreducible stoping sets in base matrix}

In this section, we will prepare several theorems 
regarding the maximal correctable burst length that are required for the argument 
in Section \ref{band splitting permutation}.

\subsection{Maximal correctable burst length of base matrices} \label{Wmax_thm}

Sridharan et. al \cite{Sridharan} studied the maximal correctable
burst length of protograph LDPC codes.  They showed a tight relationship 
between $W_{max}(B)$ and $W_{max}(H)$ where $H$ is a parity check matrix 
obtained by lifting up a base matrix $B$.
The next theorem states this relationship.

\begin{thm}[Maximal correctable burst length (\cite{Sridharan})]\label{thm:WmaxBandWmaxH}
Assume that a base matrix $B \in \Bbb F_2^{m \times n}$ is given.
Let $H$ be a parity check matrix obtained by lifting up $B$.
The following inequalities hold:
\begin{eqnarray}
(W_{max}(B)-1)M < W_{max}(H) < (W_{max}(B)+1)M.
\end{eqnarray}
\end{thm}

Theorem \ref{thm:WmaxBandWmaxH} indicates that
the maximal correctable burst length of a protograph LDPC code 
is nearly determined by $W_{max}(B)$. This means that 
an appropriate column permutation for a base matrix $B$ 
might be able to improve the maximal correctable burst length of a resulting 
photograph code. Of course, $(l,r,L)$ SC-codes belong to 
the class of protograph LDPC codes. It is reasonable to 
devise an appropriate column permutation for $B(l,r,L)$, 
which will be discussed in the next section.

\subsection{Irreducible stopping sets in $B(l,r,L)$}

The maximal correctable burst length of the base matrix $B(l,r,L)$
is determined by the set of irreducible stopping sets in $B(l,r,L)$.
In this subsection, we will show a structural property on 
the set of irreducible stopping sets in $B(l,r,L)$.

Let us denote the base matrix of the $(l,r, L)$ SC-codes as
\[
B(l,r,L)=(\bm{b}_1,\bm{b}_2,\ldots,\bm{b}_{kL})\in \mathbb{F}^{m\times kL}_2.
\]
A {\em block} $T_i (i \in [1,L])$ that is a subset of indices is defined by
\begin{equation}\label{block T}
T_i = \{ {(i-1)k+1},{(i-1)k+2},\ldots,{(i-1)k+k} \}.
\end{equation}
From the structure of $B(l,r,L)$ ({\it i.e.}, Fig. \ref{fig:lrL_base_matrix}), it is easy to 
see that  $\bm{b}_{\alpha}=\bm{b}_{\beta}$ holds if and only if ${\alpha}, {\beta} \in T_i$.
The next theorem characterizes the structure of irreducible stopping sets in $B(l,r,L)$.

\begin{thm}[Irreducible stopping sets of base matrix]\label{thm:SC_SM}
The set of irreducible stopping sets in the base matrix $B(l,r,L)$ is given by
\begin{eqnarray}
\mathcal{Q}'(B(l,r,L))=\{ \{ {\alpha}, {\beta} \} \mid {\alpha},{\beta} \in T_i, i \in [1,L]\}.
\end{eqnarray}
\end{thm}
The theorem states 
that an irreducible stopping set consists of two column indices belonging to the same block.

\begin{prf}
Suppose an ordered index set $S= (j_1,\ldots, j_u ) \subseteq [1,n]$ is given 
where $j_1 < j_2 < \cdots < j_u$.
The sub-matrix corresponding to $S$ is written as
$
B(l,r,L)_{(j_1,\ldots, j_u)}=(\bm{b}_{j_1},\bm{b}_{j_2},\ldots,\bm{b}_{j_u}).
$

We will first show a sufficient condition that $S$ is not a stopping set.
Assume that 
$
\bm{b}_{\alpha} \ne \bm{b}_{\beta}
$ 
holds for any $\alpha, \beta \in S (\alpha \neq \beta)$.
Let us focus on the first nonzero element of the first column of the sub-matrix $B(l,r,L)_{(j_1,\ldots, j_u )}$.
Due to the assumption that $\bm{b}_{\alpha} \ne \bm{b}_{\beta}$ and 
the definition of $B(l,r,L)$, it is evident that the row corresponding to 
the first nonzero element has Hamming weight 1.
This means that $S$ cannot be a stopping set in this case.

By using this sufficient condition, we can immediately 
show that any stopping set of $B(l,r,L)$ contains two different indices 
which belong to the same block. 
In other words, any stopping set must contain  $(\alpha, \beta)$ satisfying 
$
\bm{b}_{\alpha} = \bm{b}_{\beta} (\alpha \ne \beta).
$ 
If a stopping set without such a pair exists,
it contradicts the sufficient condition shown above.

It is clear that a pair of indices $(\alpha, \beta) (\alpha, \beta \in [1,kL], \alpha \ne \beta)$ is 
an irreducible stopping set if both indices $\alpha$ and $\beta$ belong to the same block.
The last job is to show that there are no irreducible stopping sets with size larger than 2.
Suppose that $S$ is an irreducible stopping set with size larger than 2. From the above argument, $S$ must contain 
at least a pair of two elements that belong to the same block. Since such a pair constitutes an irreducible stopping set,
it contradicts 
the assumption that $S$ is an irreducible stopping set. This completes the characterization of
the set of irreducible stopping sets of $B(l,r,L)$.
\qed
\end{prf}

\subsection{Burst erasure correcting capability of $(l,r,L)$ SC-codes}

An immediate application of Theorem \ref{thm:SC_SM} is to analyze the burst erasure correcting capability 
of $(l,r,L)$ SC-codes.
The size of irreducible stopping set is two and the minimal length of the stopping set is thus two;
we have $Span(B(l,r,L))=2$. 
This gives $W_{max}(B(l,r,L))=1$ and we can utilize 
Theorem \ref{thm:WmaxBandWmaxH}  to obtain lower and upper bounds on maximal correctable burst length of
$(l,r,L)$ SC-codes:
\begin{eqnarray}\label{eqn:SC_Wmax}
0 < W_{max}(H) < 2M,
\end{eqnarray}
where $H$ represents a parity check matrix of $(l,r,L)$ SC-codes.
By dividing both sides in (\ref{eqn:SC_Wmax}) by the code length $kLM$,
we have inequalities for the maximal correctable burst ratio:
\begin{eqnarray}\label{eqn:SC_lambda}
0 < \lambda_{max} < \frac{2}{kL}. 
\end{eqnarray}
It is clear that $\lambda_{max}$ converges to zero when $L$ goes to infinity.
This inequality presents that the conventional $(l,r,L)$ SC-codes have 
poor burst erasure correcting capability in the asymptotic regime when $L \rightarrow \infty$.
This result justifies the observation made by Ohashi et. al \cite{multi SC}.

\section{Band Splitting Permutations}\label{band splitting permutation}
In this section, we will propose band splitting permutations (BSP)
for the base matrix $B(l,r,L)$.
The BSP is designed to improve the span of $B(l,r,L)$.

\subsection{Definition}

\begin{figure}[t]
\begin{center}
\includegraphics[scale=0.25]{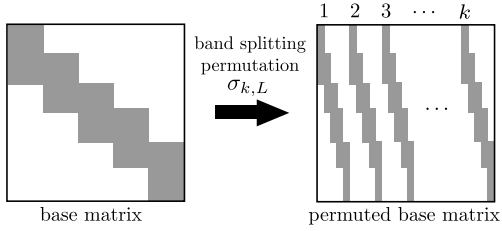}
\end{center}
\caption{The structure of conventional base matrix and permuted base matrix.}
\label{fig:divide_base}
\end{figure}

When a BSP $\sigma_{k,L}$ is applied to a base matrix $B(l,r,L)$, we have permuted base 
matrix with multiple bands as shown in Fig. \ref{fig:divide_base}.
The formal definition of BSP $\sigma_{k,L}$ is given as follows:
According to Cauchy's two-line notation on a permutation, 
the permutation $\sigma_{k,L}$ is described as 
\begin{eqnarray}\label{eqn:BSP}
\sigma_{k,L}=
\left(
\begin{array}{cccc}
1&2&\ldots&kL\\
f(1)&f(2)&\ldots&f(kL)
\end{array}
\right).
\end{eqnarray}
The second row of two line notation, {\it i.e.}, the bijective function $f$ on $[1,kL]$, is defined by
\begin{eqnarray}\label{eqn:vector a}
\arraycolsep=2pt
\left(
\begin{array}{cccc}
f(1)&f(2)&\cdots&f(kL)
\end{array}
\right)
\!=\!
\left(
\begin{array}{cccc}
\mbox{\boldmath $a$}_1&\mbox{\boldmath $a$}_2&\cdots&\mbox{\boldmath $a$}_k
\end{array}
\right), 
\end{eqnarray}
where $\bm{a}_i(i\in[1,k]$) is given by
\begin{eqnarray}\nonumber
\bm{a}_1\!&=&\!
\left(
\begin{array}{ccccc}
1   &1\!+\!k&1\!+\!2k&\cdots&1\!+\!(L\!-\!1)k
\end{array}
\right)\\\nonumber
&\vdots&\\
\label{eqn3:vector a s=k}
\bm{a}_k\!&=&\!
\left(
\begin{array}{ccccc}
k   &k\!+\!k&k\!+\!2k&\cdots&k\!+\!(L\!-\!1)k
\end{array}
\right).
\end{eqnarray}
The permutation $\sigma_{k,L}$ can be seen as a block interleaver of interleaving depth $k$.
Applying $\sigma_{k,L}$ to the base matrix $B(l, r, L)=(\bm{b}_1,\ldots, \bm{b}_{kL})$, we 
obtain a column permuted version of a base matrix 
$
B^*(l, r, L)=(\bm{b}_{f(1)},\ldots, \bm{b}_{f(kL)}).
$

For example, when $k=2, L=3$, we have
\begin{eqnarray}
\bm{a}_1=\left(\begin{array}{ccc}1&3&5\end{array}\right),\ 
\bm{a}_2=\left(\begin{array}{ccc}2&4&6\end{array}\right)\\
\sigma_{2,3}=
\left(
\begin{array}{cccccc}
1&2&3&4&5&6\\1&3&5&2&4&6
\end{array}
\right).
\end{eqnarray}
Applying $\sigma_{2,3}$ to $B(3,6,3)$, the permuted base matrix is obtained as
\begin{equation}
B^*(3,6,3)=
\left(
\begin{array}{cccccc}
1&0&0&1&0&0\\
1&1&0&1&1&0\\
1&1&1&1&1&1\\
0&1&1&0&1&1\\
0&0&1&0&0&1
\end{array}
\right).
\end{equation}

Let us define $k \times L$ matrix $A$ by
\begin{equation}\label{matrixA}
A = 
\left(
\begin{array}{c}
\bm{a}_1 \\
\vdots \\
\bm{a}_k \\
\end{array}
\right).
\end{equation}
It is easy to see that $i$-th column of $A$ corresponds the block $T_i$ 
which is defined by (\ref{block T}).
This implies that column vectors in $B(l,r,L)$ belonging to the same block
are rearranged in $B^*(l, r, L)$ as apart as possible. This property enhances 
the span of the base matrix.

\subsection{Bounds on maximal correctable burst length}

By lifting up the permuted base matrix $B^*(l,r,L)$, we can obtain 
a parity check matrix of permuted $(l,r,L)$ SC-codes.
The next theorem provides upper and lower bounds
on the maximal correctable burst length 
of permuted $(l,r,L)$ SC-codes. This is the main contribution of this work.
\begin{thm}[Bounds on maximal correctable burst length]\label{thm:BSP Wmax}
Let $B^*(l,r,L)$ be the permuted base matrix defined above.
Let $H$ be a parity check matrix obtained by 
lifting up $B^*(l,r,L)$ with the lift up factor $M$.
The maximal correctable burst length $W_{max}(H)$ of the permuted SC-code 
satisfies the following inequalities:
\begin{eqnarray}\label{eqn:BSP Wmax}
(L-1)M < W_{max}(H) < (L+1)M.
\end{eqnarray}
\end{thm}
\begin{prf}

Assume that  
$
S = \{ i_1,i_2,\ldots, i_u \} \subseteq [1,n]
$
is a stopping set of $B(l,r,L)$.
The BSP maps $S$ to 
\[
S^* = \{ f^{-1}(i_1), f^{-1}(i_2),\ldots, f^{-1}(i_u) \}.
\]
Note that $S^*$ is also a stopping set of $B^*(l,r,L)$ because 
$B^*(l,r,L)_{\{f^{-1}(i_1), f^{-1}(i_2),\ldots, f^{-1}(i_u) \}}$ contains a row of weight 1 as well.
This means that there is one-to-one correspondence between stopping sets in $B(l,r,L)$ and
those in $B^*(l,r,L)$.
Theorem \ref{thm:SC_SM} indicates that a non-empty irreducible stopping set 
consists of two indices in the same block. Assume that a pair $\alpha, \beta \in [1,kL]$ is
such a pair of indices. From a definition of the matrix $A$ in (\ref{matrixA}), it is clear that
$\alpha$ and $\beta$ belong to the same column in $A$. The definition of $f$ in (\ref{eqn:vector a}) thus leads to 
the inequality 
$
|f^{-1} (\alpha)-f^{-1} (\beta)| \ge L
$
that implies the length of irreducible stopping sets in $B^*(l,r,L)$ is larger than or equal to $L+1$.
Note that the equality holds when $\alpha$ and $\beta$ are consecutive.
From the definition of the span (\ref{spandef}), we thus have $Span(B^*(l,r,L))=L+1$
and this implies $W_{max}(B^*(l,r,L))=L$.
By using Theorem \ref{thm:WmaxBandWmaxH},  the claim of this theorem is obtained.
\qed
\end{prf}

The inequalities of Theorem \ref{thm:BSP Wmax} indicates that 
the maximal correctable burst length of the permuted $(l,r,L)$ SC-codes 
is proportional to the number of sections $L$.
The inequality (\ref{eqn:SC_Wmax})  indicates that
the maximal correctable burst length does not depend on $L$
for the case of the conventional $(l,r,L)$ SC-codes.
This result clearly shows the advantage of the permuted SC-codes 
over the conventional ({\it i.e.,} non-permuted) SC-codes with respect to 
the maximal correctable burst length.

\subsection{Maximal correctable burst ratio}
\label{s=k_lambda}
In this subsection, we focus on the maximal correctable burst ratio $\lambda_{max}$
of the permuted $(l,r,L)$ SC-codes.

By dividing both sides in (\ref{eqn:BSP Wmax}) by the code length $kLM$,
we can obtain following inequalities 
for the maximal correctable burst ratio:
\begin{eqnarray}\label{eqn:lambda}
\frac{L-1}{kL} < \lambda_{max} < \frac{L+1}{kL}.
\end{eqnarray}
From (\ref{eqn:lambda}), it is clear that 
$\lambda_{max}$ converges to $1/k$ when $L\rightarrow \infty$.
On the other hand, the design rate $R(l,r,L)$ of the $(l,r,L)$ SC-codes converges to 
$1-1/k$ as $L$ goes to infinity.
From these results,  we have
\begin{eqnarray}\label{eqn:lambda_and_R}
\lim_{L \to \infty} \left(\lambda_{max}+R(l,r,L) \right)=1
\end{eqnarray}
that indicates asymptotic optimality of permuted $(l,r,L)$ SC-codes in terms of 
burst erasure correction with erasure-BP. Note that no binary linear code of length $n$ with design rate $r$
can correct burst erasures of length larger than $n (1-r)$.

\section{Numerical results}\label{experiment}

\begin{figure}[tbp]
  \begin{center}
    \includegraphics[width=8cm,height=6cm]{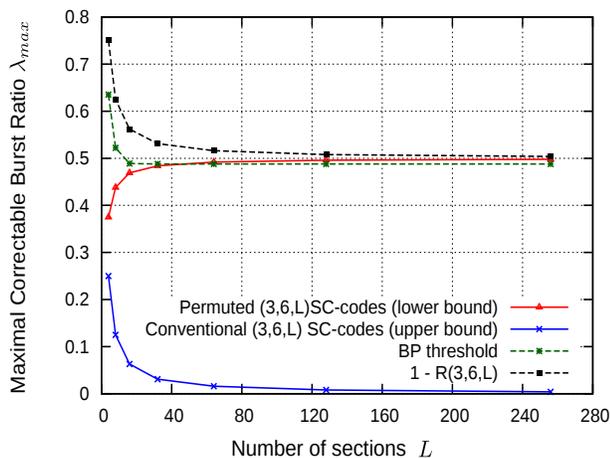}
  \end{center}
  \caption
{
Relation between $L$ and $\lambda_{max}$ of $(l,r,L)=(3,6,L)$ SC-codes
}
  \label{fig:lambda_graph1}
\end{figure}

We have seen that the maximal correctable burst ratio of  permuted SC-codes can 
be approximated by $\lambda_{max} \simeq 1/k$ when $L$ is large enough.
We will here show the relationship between  $\lambda_{max}$ and $L$ when $L$ is finite.
Figure \ref{fig:lambda_graph1} presents the bounds on $\lambda_{max}$ and 
the BP threshold of $(3,6,L)$ SC-codes.
The horizontal axis represents the number of sections $L$ 
and the vertical axis is 
the maximal correctable burst ratio $\lambda_{max}$.
It can be observed that $\lambda_{max}$ of the conventional $(3,6,L)$ SC-codes is 
decreasing as $L$ increases.
On the other hand, we can see that $\lambda_{max}$ of the permuted $(3,6,L)$ SC-codes 
converges to $1/2$. It should be remarked that 
$\lambda_{max}$ of the permuted SC-codes
is higher than the BP threshold $\theta(3,6,L)$ when $L \ge 80$.
For example, when $L=128$, $\lambda_{max}$ of the permuted SC-codes is 0.496 
but the BP threshold $\theta(3,6,128)$ is 0.488.
Assume the case where the lift up factor $M \rightarrow \infty$.
A combination of the conventional $(l,r,L)$ SC-codes and an ideal symbol interleaver
that can convert a single burst erasure into a memoryless random erasures may 
achieve the $\lambda_{max} = \theta(3,6,L)$. Thus, the permuted SC-codes 
yields better asymptotic burst erasure correcting performance when $L$ is large enough.

\begin{figure}[tbp]
  \begin{center}
    \includegraphics[width=8cm,height=6cm]{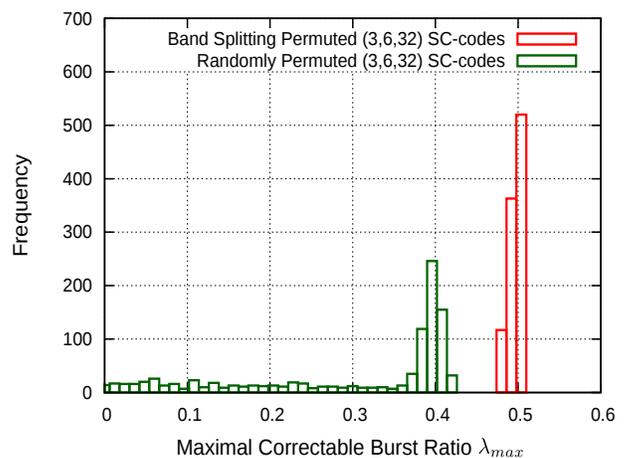}
  \end{center}
  \caption
{
Histogram of $\lambda_{max}$ of $(l,r,L)=(3,6,32)$ SC-codes $(M=40$, 1000 samples)
}
  \label{fig:histogram_graph}
\end{figure}

Figure \ref{fig:histogram_graph} presents histograms of $\lambda_{max}$ 
of randomly permuted $(3,6,32)$ SC-codes and permuted $(3,6,32)$ SC-codes ({\it i.e.}, proposed codes),
which are obtained by computer experiments with 1000 samples for each.
The lift up factor is assumed to be $M=40$.
A randomly permuted SC-code is generated as follows: 
a parity check matrix of conventional SC-codes is at first produced and 
a uniformly random column permutation is then applied to it.
It can be observed that $\lambda_{max}$'s of randomly permuted SC-codes
are far less than those of the proposed SC-codes.
This result suggests that it is not trivial to find a superior permutation 
which provides better burst erasure correcting capability than that of 
a systematically designed BSP.

%

\section*{Acknowledgement}
This work was supported by JSPS Grant-in-Aid for Scientific Research (B) Grant Number 25289114.


\begin{thebibliography}{99}
\bibitem{Low-Density Parity-Check}R. G. Gallager, ``Low-Density Parity-Check Codes,'' in Research Monograph series, Cambridge, MIT Press 1963.

\bibitem{spatially coupled code}S. Kudekar, T. Richardson, and R. Urbanke, ``Threshold saturation via spatial coupling: Why convolutional LDPC ensembles perform so well over the BEC,'' IEEE Int. Symp. Inf. Theory, pp. 684-688, Jun. 2010.

\bibitem{convolutional code}A. J. Felstrom and K. Zigangirov, ``Time-varying periodic convolutional codes with low density parity check matrix,'' IEEE Trans. Info. Theory, vol. 45, no. 6, pp. 2181-2191, Sep. 1999.

\bibitem{Sridharan1}A. Sridharan, ``Design and analysis of LDPC convolutional codes,'' Ph.D. dissertation, University of Notre Dame, Notre Dame, Indiana, 2005.

\bibitem{Sridharan2}A. Sridharan, M. Lentmaier, D. J. Costello, Jr., and K. S. Zigangirov, ``Convergence analysis of a class of LDPC convolutional codes for the erasure channel,'' in Proc. of the Allerton Conf. on Commun., Control, and Computing, Monticello, IL, USA, Oct. 2004.

\bibitem{Lentmaier}M. Lentmaier, D. V. Truhachev, and K. S. Zigangirov, ``To the theory of low-density convolutional codes. II,'' Probl. Inf. Transm. , no. 4, pp. 288-306, 2001.

\bibitem{multi SC}R. Ohashi, K. Kasai, and K. Takeuchi, ``Multi-dimensional spatially-coupled codes,'' IEEE Int. Symp. Inf. Theory, pp. 2448-2452, 2013.

\bibitem{Hosoya column permutation}G. Hosoya, H. Yagi, T. Matsushima, and S. Hirasawa, ``A modification method for constructing low-density parity-check codes for burst erasures,'' IEICE Trans. Fundamentals, vol. E89-A, no. 10, pp. 2501-2509, Oct. 2006.

\bibitem{Wadayama column permutation}T. Wadayama, ``Greedy construction of LDPC codes for burst erasure channel,'' IEICE Tech. Report, vol. 104, no. 302, pp. 35-40, Sep. 2004.

\bibitem{protograph code} J. Thorpe, ``Low-Density Parity-Check(LDPC) Codes Constructed from Protographs,'' IPN Progress Report, pp. 42-154, Aug. 2003.

\bibitem{density evolution} M. Lentmaier, A. Sridharan, D. j. Costello Jr., and K. S. Zigangirov, ``Iterative Decoding Threshold Analysis for LDPC Convolutional Codes,'' IEEE Trans. Inf. Theory, vol. 56, no. 10, pp. 5274-5289, Oct. 2010.

\bibitem{Wmax}M. Yang, and W. Ryan, ``Performance of efficiently encodable low-density parity-check codes in noise bursts on the EPR4 channel,'' IEEE Trans. Magn., vol. 40, no. 2, pp. 507-512, Mar. 2004.

\bibitem{stopping set}C. Di, D. Proietti, E. Teletar, T. Richardson, and R. Urbanke, ``Finite-length analysis of low-density parity-check codes on the binary erasure channel,'' IEEE Trans. Inf. Theory, vol. 48, pp. 1570-1579, Jun. 2002.

\bibitem{stopping set and Wmax}T. Wadayama, ``Ensemble analysis on minumum span of stopping sets,'' in Proc. Inform. Theory Applications Workshop, Feb. 2006.

\bibitem{Enrico}E. Paolini, and M. Chiani, ``Construction of near-optimum burst erasure correcting low-density parity-check codes,'' IEEE Trans. Inf. Theory, vol. 57, no. 5, pp.1320-1328, May. 2009

\bibitem{Sridharan}G. Sridharan, A. Kumarasubramanian, A. Thangaraj, and S.Bhashyam, ``Optimizing burst erasure correction of LDPC codes by interleaving,'' in Proc. IEEE Int. Symp. Inf. Theory, pp. 1143-1147, Jul. 2008

\end{thebibliography}
\end{document}